\documentstyle[12pt]{article}

\def\df #1. #2\par{\medbreak
  \noindent{{\tt {\bf Definition #1.}}\enspace}{\sl#2\par}%
  \ifdim\lastskip<\medskipamount \removelastskip\penalty55\medskip\fi}

\def\theorem #1. #2\par{\medbreak
  \noindent{\tt {\bf Theorem #1.}\enspace}{\sl#2\par}%
  \ifdim\lastskip<\medskipamount \removelastskip\penalty55\medskip\fi}

\def\lemma #1. #2\par{\medbreak
  \noindent{\tt {\bf Lemma #1.}\enspace}{\sl#2\par}%
  \ifdim\lastskip<\medskipamount \removelastskip\penalty55\medskip\fi}

\def\proof{\medbreak\noindent{\bf Proof}}

\def\la{\Lambda}

\def\D{{\cal D}}

\def\V{{\cal V}}

\def\be{\begin{equation}}
\def\ee{\end{equation}}

\begin{document}

\title{P.S. to the paper "Dirac type tensor equations with nonabelian gauge symmetries on
pseudo-Riemannian space"}

\author{N.G.Marchuk}

\maketitle

\begin{abstract}
We improve the concept of our previous paper
"Dirac type tensor equations with nonabelian gauge symmetries on
pseudo-Riemannian space" and present a new compact formula for
the tensor $B_\mu$.
\end{abstract}


\bigskip

There is Theorem 2 in \cite{Marchuk}.

\theorem 2. On the pseudo-Riemannian space $\V$ under consideration there exists
the solution
$H\in\la_1,\,I,K\in \la_2,\,B_\mu\in\la_2\top_1$ of the system of equations
\begin{eqnarray}
&&\D_\mu B_\nu-\D_\nu B_\mu+[B_\mu,B_\nu]=\frac{1}{2}C_{\mu\nu},\label{B:eqn}\\
&&\D_\mu H=0,\quad \D_\mu I=\D_\mu K=0,\label{DHIK}\\
&&H^2=1,\,I^2=K^2=-1,\,[H,I]=[H,K]=\{I,K\}=0.\label{HIK}
\end{eqnarray}

\medskip

Proving this theorem, I've presented in \cite{Marchuk} explicit formulas for $H,I,K$ via
components $g_{\mu\nu}$ of metric tensor and for $B_\mu$ via  $g_{\mu\nu}$ and its first
partial derivatives (huge formulas written in Addendum). But from those formulas  we can't 
conclude that $H,I,K,B_\mu$ are tensors (differential forms). 
Discusing the paper \cite{Marchuk} with
Prof.~ M.~ O. ~Katanaev, we arrive at understanding 
that tensors $H,I,K$ that satisfy (\ref{HIK}) must be taken as an
additional structure on the pseudo-Riemannian space $\V$.  This point of view leads to the
further development of the concept. Recently  the following theorem was proved. 

Let $\V$ be the pseudo-Riemannian space under consideration in \cite{Marchuk} and let 
differential forms $H\in\la_1,\,I,K\in \la_2$ that satisfy (\ref{HIK}) be defined on $\V$.

\theorem. If we take
\begin{eqnarray}
B_\mu &=& -\frac{3}{8}H\Upsilon_\mu H + \frac{1}{4}(I\Upsilon_\mu I+K\Upsilon_\mu K)
+ \frac{1}{8}H(I\Upsilon_\mu I+K\Upsilon_\mu K)H\nonumber\\
&&-\frac{1}{8}IKH\Upsilon_\mu H\,KI-\frac{1}{8}(KI\Upsilon_\mu I\,K+IK\Upsilon_\mu K\,I),
\label{B}
\end{eqnarray}
then $H,I,K,B_\mu$ satisfy equalities (\ref{B:eqn},\ref{DHIK}), where 
$\D_\mu=\Upsilon_\mu - [B_\mu,\,\cdot\,]$. \par

\proof\,\, is by direct calculation using relations 
$$
\{\Upsilon_\mu H,H\}=\{\Upsilon_\mu I,I\}=\{\Upsilon_\mu K,K\}=0,\quad
\{\Upsilon_\mu K,I\}=-\{\Upsilon_\mu I,K\},
$$
$$
[\Upsilon_\mu H,I]=[\Upsilon_\mu I,H],\quad [\Upsilon_\mu H,K]=[\Upsilon_\mu K,H],
$$
which follows from the equalities (\ref{HIK}). It can be easily seen from the formula
(\ref{B}) that $B_\mu$ is a tensor, i.e., $B_\mu\in\la_2\top_1$. Therefore we arrive at
tensors $H\in\la_1,\,I,K\in \la_2,\,B_\mu\in\la_2\top_1$ that satisfy (\ref{B:eqn}-\ref{HIK})
and we have the simple formula (\ref{B}) for $B_\mu$. 

If we replace Theorem 2 of \cite{Marchuk} by the above Theorem, then all results of 
\cite{Marchuk} are valid.

\end{document}